\documentclass[twocolumn,amsmath,amssymb, twocolumn]{revtex4}

\usepackage{graphicx}
\usepackage{dcolumn}
\usepackage{bm}
\usepackage{float}
\usepackage{amsfonts}

\begin{document}

\preprint{APS/XYZ}

\title{A new structural model for the Si(331)-(12$\times$1) reconstruction}

\author{Corsin Battaglia}
\affiliation{Institut de Physique, Universit\'e de Neuch\^atel,
2000 Neuch\^atel, Switzerland}%
\email{corsin.battaglia@unine.ch}
\homepage{http://www.unine.ch/phys/spectro}
\author{Katalin Ga\'{a}l-Nagy}
\affiliation{Dipartimento di Fisica and European Theoretical
Spectroscopy Facility (ETSF), Universit\`a di Milano, 20133 Milano,
Italy}
\author{Claude Monney}
\affiliation{Institut de Physique, Universit\'e de Neuch\^atel,
2000 Neuch\^atel, Switzerland}%
\author{Cl\'ement Didiot}
\affiliation{Institut de Physique, Universit\'e de Neuch\^atel,
2000 Neuch\^atel, Switzerland}%
\author{Eike Fabian Schwier}
\affiliation{Institut de Physique, Universit\'e de Neuch\^atel,
2000 Neuch\^atel, Switzerland}%
\author{Michael Gunnar Garnier}
\affiliation{Institut de Physique, Universit\'e de Neuch\^atel,
2000 Neuch\^atel, Switzerland}%
\author{Giovanni Onida}
\affiliation{Dipartimento di Fisica and European Theoretical
Spectroscopy Facility (ETSF), Universit\`a di Milano, 20133 Milano,
Italy}
\author{Philipp Aebi}
\affiliation{Institut de Physique, Universit\'e de Neuch\^atel,
2000 Neuch\^atel, Switzerland}%

\date{\today}

\begin{abstract}
A new structural model for the Si(331)-(12$\times$1) reconstruction
is proposed. Based on scanning tunneling microscopy images of
unprecedented resolution, low-energy electron diffraction data, and
first-principles total-energy calculations, we demonstrate that the
reconstructed Si(331) surface shares the same elementary building
blocks as the Si(110)-(16$\times$2) surface, establishing the
pentamer as a universal building block for complex silicon surface
reconstructions.
\end{abstract}

 \keywords{}

\maketitle

The study of semiconductor surface reconstructions has been an area
of active research for many years and has gained tremendous
importance with the advent of low-dimensional heteroepitaxial
semiconductor nano\-structures such as quantum dots and quantum
wires \cite{Ross99}. The creation of a surface results in broken
bonds, called dangling bonds. Dangling bonds are energetically
unfavorable causing surface atoms to rearrange or reconstruct. This
often results in highly complex atomic structures, whose
determination remains a formidable challenge and requires the
complementary role of different experimental
and theoretical methods.\\
In order to lower the surface energy, silicon surfaces adopt a
variety of strategies allowing to reduce the number of dangling
bonds. Despite the large number of known surface reconstructions,
one frequently encounters common elementary structural building
blocks \cite{Battaglia08b,Battaglia07}. Identifying these building
blocks is important not only for a better understanding of these
surfaces, but also serves as a guide for the elaboration of new
structural models.\\
Two of the most important strategies, encountered for instance on
Si(100) \cite{Chadi79} and Si(111) \cite{Takayanagi85}, are
respectively the formation of dimers, where two surface atoms pair
up to eliminate their dangling bonds, and the appearance of adatoms,
which bond to three surface atoms thus saturating three dangling
bonds. An important step towards the understanding of high-index
group IV surfaces with a surface normal in between the (111) and
(100) direction was the introduction of an additional reconstruction
element by Dabrowski \textit{et al.} \cite{Dabrowski95}. They
proposed a six-fold coordinated surface self-interstitial which is
captured by a conglomerate of surface atoms
\cite{Laracuente98,Stekolnikov03}. This concept was subsequently
adapted by An \cite{An00} and theoretically analyzed by Stekolnikov
\cite{Stekolnikov04,Stekolnikov04b} to explain the pairs of
pentagons observed in scanning tunneling microscopy (STM) images of
the reconstructed
Si(110) surface.\\
\begin{figure}
\centering
\includegraphics{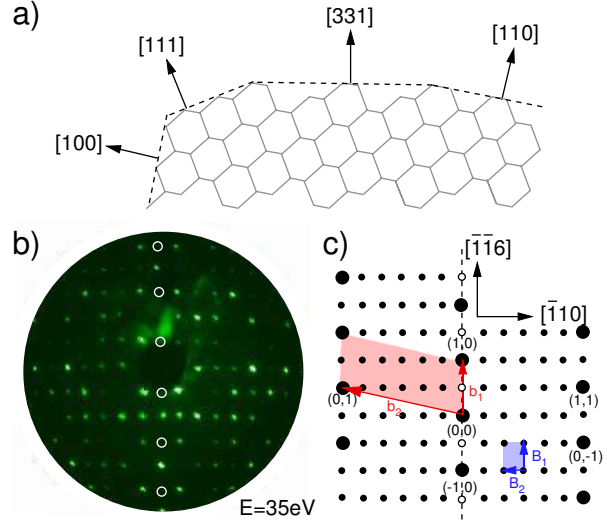}%
\caption{\label{fig:LEED}(Color online) a) Side cutaway showing the
crystal lattice of silicon in the $(\bar{1}10)$ plane. The dashed
line follows the bulk-terminated surface for several important
orientations. b) Experimental LEED pattern at 35 eV beam energy. The
positions of missing spots are indicated by the white circles. c)
Sketch of the LEED pattern with (1$\times$1) (red) and (12$\times$1)
(blue) reciprocal unit cells and spot labels (black). The bulk
directions are also given. The positions of the missing spots are
indicated by empty black circles. The orientation of the glide plane
is indicated by a dashed line.}
\end{figure}
\noindent In this letter we focus on the atomic structure of the
Si(331)-(12$\times$1) reconstruction. We present high-resolution STM
images resolving for the first time rows of pentagons very similar
to the ones observed on Si(110)-(16$\times$2). Si(331), whose
surface normal is located 22.0$^o$ away from the (111) direction
towards (110) (see Fig. \ref{fig:LEED}a)), is an important surface,
since it is the only confirmed planar silicon surface with a stable
reconstruction located between (111) and (110). Since the discovery
of the Si(331)-(12$\times$1) reconstruction more than 17 years ago
\cite{Wei91} several structural models containing dimers and adatoms
have been proposed \cite{Olshanetsky98,Gai01}. However, none of
these models is able to explain the pentagons observed in our STM
images. Combining the complementary strength of STM, low-energy
electron diffraction (LEED), and first-principles total-energy
calculations, we derive a new structural model containing surface
self-interstitials as basic
building blocks.\\
Sample preparation and experiments were carried out in a ultra-high
vacuum chamber with a residual gas pressure below $3\times10^{-11}$
mbar equipped with an Omicron LT-STM and Omicron Spectaleed
LEED/Auger optics. Boron doped Si(331) samples from CrysTec with a
resistivity of 0.1-30 $\Omega$ cm were slowly degassed by direct
current heating up to 1060$^o$C while keeping the pressure below
$5\times10^{-10}$ mbar, followed by subsequent cleaning via repeated
flashing to 1260$^o$C and slow cooling across the (1$\times$1) to
(12$\times$1) phase transition at 810$^o$C \cite{Wei91}. This
procedure gives an almost perfectly ordered surface with a very low
number of defects. Special care was taken in order to avoid Ni
contamination. STM measurements were performed at 77 K using etched
W tips.\\
\begin{figure}\centering
\includegraphics{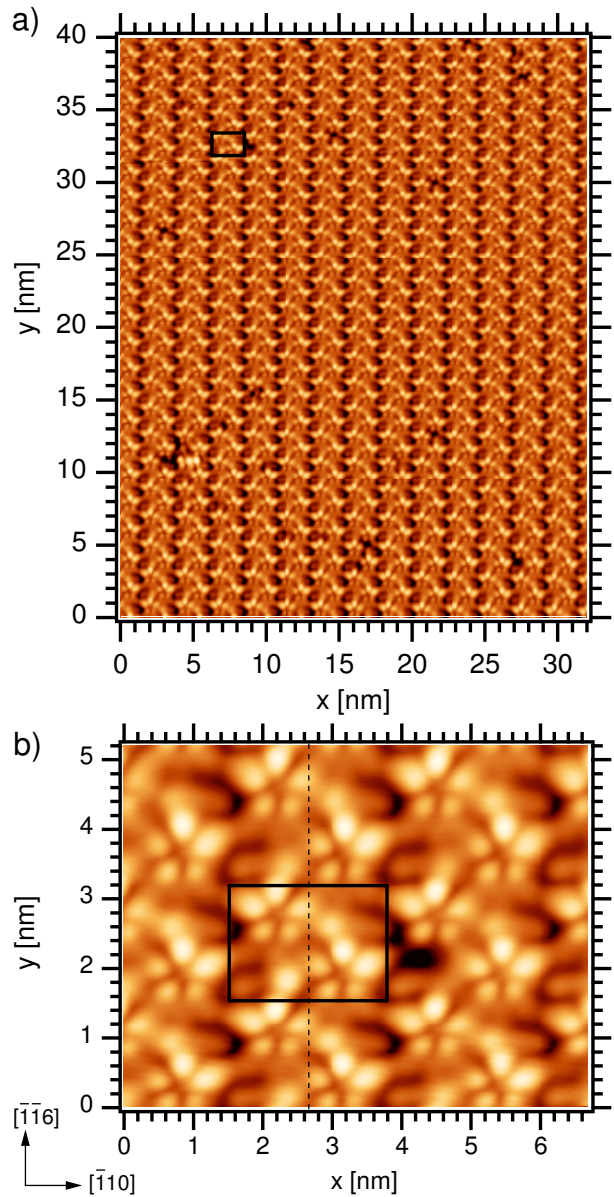}%
\caption{\label{fig:STM331}(Color online) a) Large scale STM
topography of the annealed Si(331) surface. b) High-resolution image
with unit cell (full line) and the glide plane (dashed line)
indicated. Bias voltage 2.0 V, set-point current 0.06 nA. }
\end{figure}
\noindent The theoretical results are based on first-principles
calculations performed with ABINIT \cite{abinit}. This code employs
a density-functional theory scheme within a plane-wave expansion,
where we have used norm-conserving pseudopotentials and the
local-density approximation for the exchange-correlation potential.
ABINIT has already been successfully applied to describe the Si(113)
surface \cite{GaalNagy07}. The present (331) surface has been
studied using a slab supercell containing 240 bulk silicon atoms,
plus a variable number of surface atoms (depending on the
reconstruction). The bottom surface was passivated with 36 hydrogen
atoms and adjacent slabs are separated by 9.4 \AA\; of vacuum.
Convergence required a kinetic-energy cutoff of 218 eV and a single
(nonzero) k-point in the Brillouin zone, yielding an accuracy of the
calculated forces of 0.077 eV/\AA. In order to accelerate the
convergence of the self-consistent cycle we introduced a metallic
broadening of 0.054 eV.
\\
Figure \ref{fig:LEED}b) presents a normal incidence LEED pattern of
the Si(331)-(12$\times$1) reconstruction. The corresponding sketch
containing reciprocal lattice vectors and spot labels $(h,k)$ is
shown in Fig. \ref{fig:LEED}c). The bulk-terminated surface is
chosen as the reference for indexing. The reciprocal unit cell
vectors $\mathbf{B}_1$ and $\mathbf{B}_2$ of the reconstructed
surface can be expressed using the reciprocal unit cell vectors
$\mathbf{b}_1$ and $\mathbf{b}_2$ of the bulk-terminated (331)
surface
$$
\left(
  \begin{array}{c}
    \mathbf{b_1} \\
    \mathbf{b_2} \\
  \end{array}
\right)=\left(
          \begin{array}{cc}
            2 & 0 \\
            1 & 6 \\
          \end{array}
        \right)
        \left(
          \begin{array}{c}
            \mathbf{B_1} \\
            \mathbf{B_2} \\
          \end{array}
        \right).
$$
However, since in general we find 11 satellite diffraction spots in
between the integer spots along the $[\bar{1}10]$ direction, the
reconstruction
is conventionally called the (12$\times$1) reconstruction in the literature.\\
The spot intensities in the LEED pattern in Fig. \ref{fig:LEED}b)
exhibit a mirror symmetry along the $[\bar{1}\bar{1}6]$ direction.
Furthermore, a careful analysis of the LEED spot intensities as a
function of energy (not shown) reveals that systematically
\textit{all} half-order spots $(\pm n h/2,0)$, $n$ being an odd
integer, are missing for \textit{all} beam energies (see empty
circles in Fig. \ref{fig:LEED}b) and c). Although spot intensities
vary as a function of energy and even vanish at some energies due to
diffraction effects, the $(\pm n h/2,0)$ spots exhibit no intensity
at \textit{all} beam energies. Such missing spots indicate the
existence of a glide plane along $[\bar{1}\bar{1}6]$ in real space
\cite{Holland73,Yang83}. This glide plane also implies the mirror
symmetry observed for the LEED intensities in reciprocal space
\cite{Debe77}. Thus LEED firmly establishes the
presence of a glide plane in the structure.\\
We now turn to our STM data. The large scale topography image
presented in Fig. \ref{fig:STM331}a) shows the high quality of the
surface. Besides the perfect long range order only a few local
defects are present on the surface. As already observed by several
other groups \cite{Tanaka92, Hibino93, Tanaka94,Hibino96, Hibino96b,
Gai01}, the Si(331)-(12$\times$1) reconstruction consists of similar
mounds arranged into zigzag chains running along the
$[\bar{1}\bar{1}6]$ direction separated by trenches. Focusing now on
the high-resolution image in Fig. \ref{fig:STM331}b) we see that
each of the mounds consists of five protrusions forming a pentagon.
A further protrusion may be identified linking two successive
pentagons within the same chain.  Pentagons with the same dimensions
were already observed on Ge(110)-c(8$\times$10)
\cite{Ichikawa95,Gai98} as well as on Ge(110)-(16$\times$2)
\cite{Ichikawa95} and on Si(110)-(16$\times$2) \cite{An00}. To our
knowledge this is {\em the first observation of such pentagons on a
surface away from the (110) orientation}, indicating their high
stability and confirming their fundamental role as an elementary
building block in semiconductor
surface reconstructions.\\
Inspired by structural elements encountered on reconstructed Si(113)
and Ge(113) surfaces \cite{Dabrowski95,Laracuente98}, An \textit{et
al.} \cite{An00} have proposed an adatom-tetramer-interstitial (ATI)
model for the pentagons observed on the (110) surfaces. Its
stability has subsequently been tested theoretically by means of
first-principles total-energy calculations
\cite{Stekolnikov04b,Stekolnikov04}. \\
In the following, we develop a coherent structural model for the
Si(331)-(12$\times$1) reconstruction inspired by the ATI model and
discuss similarities and differences with the (110) case. In a first
step we need to determine the registry of the surface reconstruction
with respect to the bulk. This is not a trivial task, so we proceed
in two steps. We focus first on the position of the surface
reconstruction with respect to the underlying bulk along the
$[\bar{1}10]$ direction. Here the occurrence of the glide plane
symmetry gives us the clue. Inspection of Fig. \ref{fig:STM331}b)
shows that a glide plane is found at the center of the zigzag chain
(dashed line) consistent with the observation of missing spots in
the LEED pattern. The glide plane found on the surface must also be
a glide plane of the bulk, since the space group of the bulk
contains all symmetry elements of the surface. A side view of the
bulk terminated Si(331) surface is shown in Fig.
\ref{fig:Model331}a) with a top view of our model in Fig. 3b). The
dashed line represents the glide plane. Fig. \ref{fig:Model331}c)
offers a graphical proof for the existence of the glide plane in the
bulk. After mirror reflection along the glide plane line, a
translation by half the unit vector $-\mathbf{A}_1/2$ is necessary
to obtain the original registry.
\\
\begin{figure}[th!]
\centering
\includegraphics{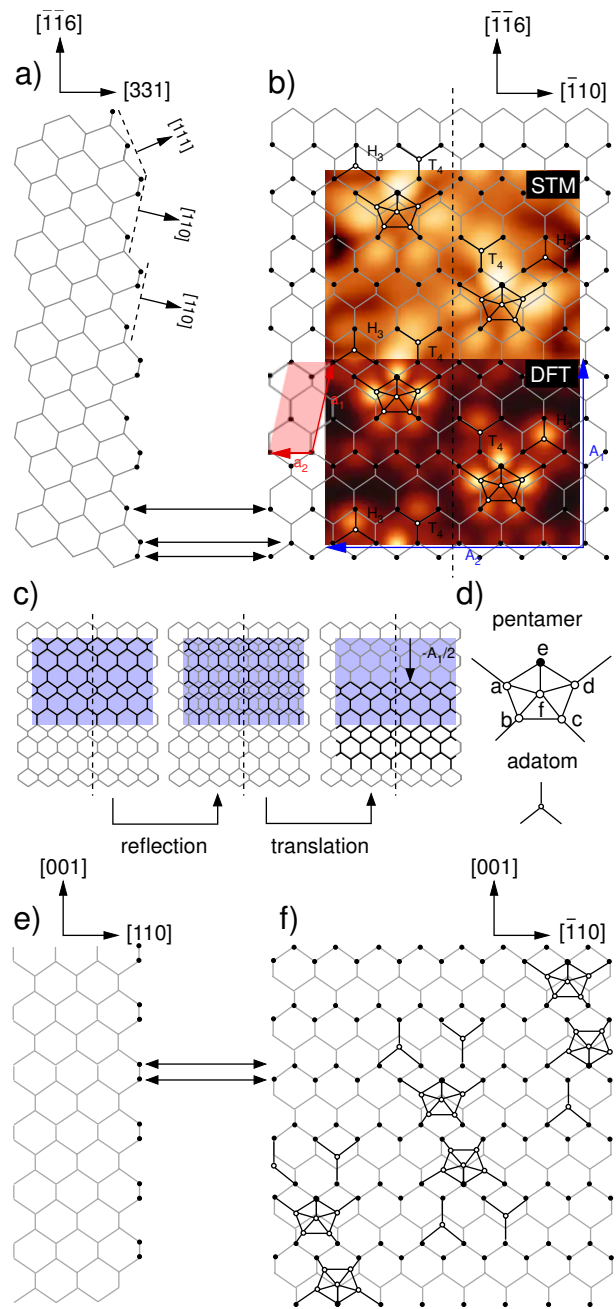}%
\caption{\label{fig:Model331}(Color online) a) Side  and b) top view
of the Si(331) surface.  Black dots indicate the position of
dangling bonds, some of which are saturated by pentamers and adatoms
(empty circles). The (1$\times$1) (red) and (12$\times$1) (blue)
unit cells are shown. For comparison an experimental and a simulated
STM image is underlaid. The dashed line represents the glide plane.
c) Graphical proof for the existence of the glide plane. d) Sketch
of the pentamer and the adatom with atom labels. e) Side and f) top
view of the Si(110) surface. (see text)}
\end{figure}
\noindent After determining the registry of the surface
reconstruction with respect to the $[\bar{1}10]$  direction  we now
need to study the registry with respect to the $[\bar{1}\bar{1}6]$
direction. Here we benefit from a comparison with the Si(110)
surface shown schematically in Fig. \ref{fig:Model331}e) and f). For
a sketch of the complete model for the Si(110)-(16$\times$2)
reconstruction including the steps along $[\bar{1}12]$ see Ref.
\cite{Stekolnikov04b}. According to the ATI model for the Si(110)
surface, the pentagon seen in STM images consists of four adatoms
($a$,$b$,$c$,$d$, empty circles) forming the tetramer and one
surface atom ($e$, black dot) belonging to the first atomic layer
(see Fig. \ref{fig:Model331}d)). The six-fold coordinated
interstitial atom ($f$, empty circle) is located at the center of
the pentagon slightly below the tetramer plane and consequently not
directly visible in STM images (see simulated STM images in Ref.
\cite{Stekolnikov04b}). The resulting structural element formed by
the tetramer, atom $e$ and $f$ combined is called a pentamer
\cite{Stekolnikov04b}.
\\
In order to integrate the pentamer building block into our model we
note that the arrangement of dangling bonds on the bulk-truncated
surface represented by black dots and marked by the double headed
arrows in Fig. \ref{fig:Model331} differs between the Si(110) and
Si(331) surface. Whereas dangling bonds on the Si(110) surface occur
in double rows running along $[\bar{1}10]$, double rows alternate
with single rows of dangling bonds on the bulk-terminated Si(331)
surface. For the Si(110) structure, atom $e$ actually belongs to one
of these dangling bond double rows. Consequently we also anchor the
two pentamers per unit cell  required by STM on the double dangling
bond row of the bulk-terminated Si(331) surface. It is important to
note that anchoring the pentamers in this way provides exactly the
same local binding configuration as on the Si(110) surface, since
careful comparison of the sideviews in Fig. \ref{fig:Model331}a) and
e) reveals that the bulk-truncated Si(331) surface can be viewed as
a highly stepped Si(110) surface. The selected position of the
pentagons agrees with the position observed in the STM images (see
STM image behind the model in Fig. \ref{fig:Model331}b)). Note also
that the single dangling bond row of the Si(331) surface is slightly
lower-lying than the double dangling bond row, causing an
inclination of the pentamer, in agreement with the lower intensity
of the lobes associated with adatom $b$ and $c$ in the experimental
STM
image.\\
Each pentamer saturates five of the surrounding dangling bonds
\cite{Stekolnikov04}. By introducing two pentamers per
Si(331)-(12$\times$1) unit cell, the number of dangling bonds has
been reduced from 36 to 26. Some of the remaining dangling bonds are
saturated by simple adatoms as in the case of the Si(110) surface.
%In order to find the energetically most favorable adatom
%configuration for our structural model we have determined
%theoretically the equilibrium geometries and relative surface
%energies for a number of different adatom arrangements. Details on
%the investigated models and calculations will be published elsewhere
%\cite{Battaglia08}.
In the STM image in Fig. \ref{fig:STM331}b) the
additional protrusion linking two successive pentamers indicates the
location of a first adatom, labeled $T_4$ in Fig.
\ref{fig:Model331}b) in analogy with the convention used to label
this adatom position on the (111) surface. The local binding
configuration for this $T_4$ adatom on the Si(331) surface is
exactly the same as on the (111) surface, where adatoms are common
structural building blocks. A further adatom
labeled $H_3$ saturates three more dangling bonds.\\
Introducing two $T_4$ and two $H_3$ adatoms per unit cell into our
structural model further reduces the number of dangling bonds from
26 to 14. For Si(110) Stekolnikov \textit{et al.}
\cite{Stekolnikov04b} have noted that it is energetically more
favorable to leave some surface atoms, so called rest atoms,
unsaturated than to introduce the maximum number of adatoms into the
model, since this allows a further reduction of the surface energy
by electron transfer from the adatom to the rest atom in analogy
with
the Si(111)-(7$\times$7) case \cite{Meade89}.\\
We have explicitly verified the stability of the reconstruction
containing two $T_4$ and two $H_3$ adatoms by means of
first-principles calculations testing various alternative adatom
configurations, among which the present one yields the highest
stability (details on the theoretical investigations will be
published elsewhere \cite{Battaglia08}). In Fig.
\ref{fig:Model331}b) we also show a simulated STM image based on
relaxed coordinates of our structural model, obtained by integrating
the local state density over a 2.0 eV energy window corresponding to
the experimental STM bias voltage. The DFT image is in excellent
agreement with the experimental STM image. The theoretical image
reproduces well the strong intensity of the lobes associated with
atoms $a$, $e$ and $d$ and the weaker intensity associated with
atoms $b$ and $c$ of the pentamer. In general the topographical
features in the simulated image show better resolution and less
diffuse character than the corresponding features observed in STM,
since the simulation neglects any role of the tip structure in
degrading image resolution \cite{Baski95}. In the experimental as
well as in the simulated image, individual pentagons do not exhibit
a mirror symmetry along the $[\bar{1}\bar{1}6]$ direction indicating
a distortion of the pentagon which allows to reduce internal strain.
Furthermore the $T_4$ and $H_3$ adatoms in the simulated image are
seen as a marked protrusion in agreement
with experiment. \\
It is interesting to note that on the Si(110) surfaces, pentamers
always occur in twin pairs rotated by 180$^o$ with respect to each
other, whereas on Si(331) all pentamers point into the same
direction and are further apart from each other. Comparing the
transition temperature of 810$^o$C \cite{Wei91} for
Si(331)-(12$\times$1) with 730$^o$C \cite{Yamada07} for
Si(110)-(16$\times$2), it appears that Si(331) is the more favorable
surface for pentamer formation (for comparison the transition
temperature of 870$^o$C \cite{Viernow98} for the
Si(111)-(7$\times$7) reconstruction is slightly higher). Further
detailed total-energy calculations are required to quantify the
trade-off between surface dangling bond reduction and induced
surface stress and to compare the differences in the bonding
configuration between the Si(331) and the Si(110) surface.\\
In summary, by combining STM, LEED, first-principles calculations
and by comparing similarities and differences between the
Si(331)-(12$\times$1) and Si(110)-(16$\times$2) reconstructions, we
have derived a complete structural model for the Si(331) surface
containing the pentamer as an essential ingredient. Thus besides
adatoms, dimers and tetramers, pentamers emerge as a universal
building blocks for silicon surface reconstructions.\\
Stimulating discussions with Antje Schmalstieg, Georg Held, Pascal
Ruffieux, and Oliver Gr\"{o}ning are gratefully acknowledged.
Skillfull technical assistance was provided by our workshop and
electric engineering team. This work was supported by the Fonds
National Suisse pour la Recherche Scientifique through Div. II, the
Swiss National Center of Competence in Research MaNEP, and by the
EU's 6th Framework Programme through the NANOQUANTA Network of
Excellence (NMP-4-CT-2004-500198).

%%Dont forget to show single trench defects and varius bias images
%%Discuss stability of Si(110) vs Si(331) via transition temperature and long range order/defects

\bibliography{Si331}

\end{document}